\documentclass[epsfig,12pt]{article}
\usepackage{epsfig}
\usepackage{graphicx}

\usepackage{latexsym}
\usepackage{amsmath}
\usepackage{amssymb}
\usepackage{relsize}
\usepackage{geometry}
\def\beq{\begin{equation}}
\def\eeq{\end{equation}}
\def\beqn{\begin{eqnarray}}
\def\eeqn{\end{eqnarray}}

\newcommand{\gsim}{\lower.7ex\hbox{$
\;\stackrel{\textstyle>}{\sim}\;$}}
\newcommand{\lsim}{\lower.7ex\hbox{$
\;\stackrel{\textstyle<}{\sim}\;$}}

\def\slashed#1{\setbox0=\hbox{$#1$}             
   \dimen0=\wd0                                 
   \setbox1=\hbox{/} \dimen1=\wd1               
   \ifdim\dimen0>\dimen1                        
      \rlap{\hbox to \dimen0{\hfil/\hfil}}      
      #1                                        
   \else                                        
      \rlap{\hbox to \dimen1{\hfil$#1$\hfil}}   
      /                                         
   \fi}                                        %

\begin{document}


\begin{titlepage}

\begin{flushright}
ITEP-TH-05/11
\end{flushright}

\vspace{1cm}

\begin{center}
{  \Large \bf  Angular Momentum  and
Gravimagnetization of the ${\cal N}=2$ SYM vacuum}
\end{center}
\vspace{5mm}
\begin{center}
{  \bf Alexander Gorsky}
\end{center}
\vspace{0.8mm}
\begin{center}
{\it Theory Department, ITEP, Moscow, Russia}
\end{center}

\begin{center}

{\large \bf Abstract}
\vspace {5mm}
\end{center}

In this note we discuss the gravimagnetization
of the ${\cal N}=2$ SYM vacuum in the $\Omega$-background. It is
argued that the Seiberg-Witten prepotential is related
to the vacuum density of the angular momentum in the Euclidean $R^4$ space.
The possible
role of the dyonic instantons  as the microscopic
angular momentum carriers which could yield the spontaneous
vacuum gravimagnetization is conjectured. We interpret the
dyonic instanton as a kind of the Euclidean bounce in $R^4$ similar
to one responsible for the Schwinger pair creation. The induced
angular momentum in $R^4$ is also briefly considered  in the
dual  Liouville formulation of $SU(2)$ theory via AGT relation.

\end{titlepage}

\section{Introduction}
The common strategy while investigating   the properties of a
complicated system is to put it in the proper background
fields and look at response. The background usually plays two-fold role,
first  it provides the  regularization of the possible divergences and
secondly it helps to extract the effective degrees of freedom via the
response functions.
For example, one can investigate the magnetization $<M>= \frac{d log Z}{dB}$
in the external  magnetic field $B$ or some transport coefficients like
conductivity in the external electric field.
Such strategy has been applied to the analysis of the ground state
of ${\cal N}=2$ SYM theory in \cite{nek}.
Formally the Nekrasov partition function counts the instanton contributions
to the holomorphic prepotential in the $\Omega$-deformed ${\cal N}=2$ SYM theory \cite{nek}.
The background is provided by the graviphoton field which
has managed to cure the singularity in the integration over
the instanton moduli space.

In spite of being the effective tool to get the
explicit answer the localization
approach is a little bit formal and can not
immediately provide the
answers on more "naive" questions. In particular it is
interesting to know if  instantons are organized in some effective
degrees of freedom of higher dimensions. On the other hand
there is the old suspicion that  instanton itself can be considered
as a kind of composite (see \cite{tonco} and references therein
for the recent discussion).

In this note we focus on the second aspect of the background field analysis
namely the response functions of the system. The simplest response function
corresponds to the derivative of the Nekrasov partition function with respect
to the graviphoton field.
By the construction the parameters of the $\Omega$-background play the role of the
chemical potentials for the rotations in the four-dimensional Euclidean
space-time as well as the R-symmetry rotations. Hence we are in position
to get the corresponding vev of the angular momentum in $R^4$ just differentiating
the $log Z_{Nek}$ with respect to the graviphoton field that is
from the  gravimagnetization of the ${\cal N}=2$ SYM vacuum.

Since we would like to analyze the properties of the instanton matter
it is convenient to have in mind the $D=5$ picture where the instantons
are the particles charged with respect to the external graviphoton field.
The $D=4$ prepotential is interpreted  as  particular
limit of  generalized index
in the $D=5$ SYM theory on $R^4\times S^1$ and the very Seiberg-Witten
prepotential  \cite{sw} can be
derived from the one-loop calculation with all particles in the
loop including instantons  are taken into account \cite{ln}.

The simple semiclassical limit of the partition function provides us with
a kind of surprise since it can be recognized immediately that the density
of the angular momentum diverges at arbitrarily weak graviphoton  field. This looks
like an absolute instability of the ${\cal N}=2$ SYM matter with respect to the generation
of the angular momentum in the external graviphoton field very much similar
to absolute instability of the QED matter in the external electric field
due to the Schwinger pair creation.
The $D=5$ one-loop derivation of the prepotential suggests that the analogy
with the QED in the external field can be considered more carefully.
The QED example provides us also with the toy model for the
angular momentum generation.
We shall compare QED and SYM phenomena in the external fields
using the Euclidean interpretation of the pair production process
via the bounce solution.

Having in mind the non-vanishing vev of the angular momentum
it is natural
to question on the microscopic states which are its carriers
in the Euclidean space-time. The natural particle charged with
respect to the graviphoton is D0 brane. When the graviphoton
field is switched on the stabilization of the Euclidean bounce
is due to the topological D0 charge dissolved
on the D2 brane worldvolume.
Moreover since we handle with the instanton matter
the intrinsic object carrying
the angular momentum is the dyonic instanton \cite{dyonic,town,dyonic2,emparan}
that is the circular D2 brane with the electric and topological
charges. This solution exists even without the graviphoton
background and gets stabilized by the angular momentum of the fields
proportional to the product of the instanton and electric charges.
Hence generically the phenomena of the spontaneous
gravimagnetization of the instanton matter could happen.

Since the Nekrasov partition function is identified with the Liouville
conformal block \cite{agt} we could question the issue of the angular
momentum in the Liouville theory as well. The parameters of the background
enter the expression for the conformal weights  and
the central charge. That is we shall get the interesting interpretation
for the  derivatives of the conformal block with respect to the
conformal weights of the operators involved as the angular momentum
in the physical space. We shall also comment on the additional angular momentum 
due to the surface operators. As a byproduct remark we  also 
relate the one-loop calculation in the external
field with the auxiliary problem in the AdS space which provides us
with the objects similar to conformal blocks in the Liouville theory.

The note is organized as follows. In the Section 2 we discuss the induced
momentum  in the $\Omega$-deformed ${\cal N}=2$ SYM theory and compare it with
the Euclidean description of the pair production in QED
in external field in Section 3. Section 4
concerns the possible role of the dyonic instantons in
the spontaneous gravimagnetization.
Section 5 is devoted
to the discussion of the angular momentum at the Liouville side
while in the last Section  some discussion can be found.

 \section{Angular momentum from Nekrasov partition function}

 Let us remind the main features of the Nekrasov partition function. One
 considers the ${\cal N}=2$ SYM theory and introduces the $\Omega$-background
 switching on the graviphoton RR field in $R^4$. The most simple way to describe
 $\Omega$-background is to start with the six- dimensional space-time
 and introduce $R^4$ bundle with the nontrivial $SO(4)$ connection
 over the two-dimensional torus.
 The corresponding metric reads as
 \beq
 ds^2=A dzd\bar{z} + g_{ij}(dx^i + V^i dz + \bar{V}^i d\bar{z})
 (dx^j + V^j dz + \bar{V}^j d\bar{z})
 \label{metric}
 \eeq
 where $V^i= \Omega^i_{j}x^j$ and $A$ is the torus area. The components of the
 graviphoton field strength tensor $\Omega^i_{j}$,  $\epsilon_{1,2}$
 provide the regularization of the integration over the instanton moduli space.
 The alternative description of the
 $\Omega$-background is recently suggested in \cite{nw}.

 The localization of the integral over the instanton moduli space
 with respect to the natural torus action yields the exact partition
 function of the ${\cal N}=2$ SYM in the $\Omega$ background \cite{nek}.
 It can be interpreted in terms of the weighted
 equivariant volumes of the instanton moduli spaces ${\cal M}_N$.
 \beq
 Z_{Nek}(a,\epsilon_1,\epsilon_2)= \sum_{N} q^{N} \int_{{\cal M}_N} "1"
 \eeq
 It is important that the conventional Seiberg-Witten prepotential $\cal{F}$ \cite{sw} can be derived
 at  the semiclassical limit $\epsilon_{1,2}\rightarrow 0$
 \beq
 Z_{Nek}\propto \exp(\frac{\cal{F}}{\epsilon_1 \epsilon_2}+...)
 \eeq
Note that in the $\Omega$ background the effective volume
of the system is
\beq
Vol_{eff}= \frac{1}{\epsilon_1 \epsilon_2}
\eeq
hence to some extend in the semiclassical limit 
the prepotential can be treated as the "action density"
of the vacuum state.

In what follows it will be useful for our purposes to consider the different
realization of the Nekrasov partition function as the generalized
index in the $D=5$ SYM theory \cite{nek}
\beq
Z_{Nek}= Tr_{H}(-1)^{2(J_L+ J_R)}(\exp((\epsilon_1 -\epsilon_2)J^3_{L} +
(\epsilon_1 + \epsilon_2)J^3_{R} + (\epsilon_1 + \epsilon_2)J^3_{I} +\beta H)
\label{index}
\eeq
where $J_L, J_R$ are the generators of independent rotations in  $R^4$
while $J_I$ corresponds to the R-symmetry rotation.
This representation clearly shows  that the parameters of the
deformation play the role of the generating parameters for
the angular momenta and R- symmetry rotation.

This representation fits with the derivation of the Seiberg-Witten
prepotential from the theory on $R^4\times S^1$ \cite{ln} which
involves explicit one-loop calculation where all massive  BPS
particles with the topological and electric charges
are summed over.
Since instantons are particles in five dimensions they are
treated as degrees of freedom propagating in the loop. Upon taking the proper
limit one-loop calculation in $D=5$ yields precisely the instanton Seiberg-Witten
prepotential \cite{ln}. The representation of $Z_{Nek}$
as the generalized index fits with this one-loop calculation
in the $\Omega$ background. Somewhat similar representation
of the generalized indexes has been also explored in \cite{vafa}.

Given the index representation it is evident
that  the derivatives of $log Z$ with respect to the equivariant parameters
yield the vev of the physical angular momentum $<J_{L.R}>$.
The arguments providing the similar interpretation in the
$D=4$ framework are based on the particular form of the metric
(\ref{metric}) in
the $\Omega$ background. Indeed the four-dimensional part
of the metric is deformed by the terms proportional
to the coordinates $x_{\nu}$. The response of the Lagrangian
on this deformation canonically yields the four-dimensional
angular momentum hence the $D=4$ and $D=5$ pictures are consistent
with each other.

It is convenient to extract the pure angular momentum without
the admixture of the R-rotation. To this aim one can consider the
difference $\partial_{\epsilon_1} - \partial_{\epsilon_2}$

\beq
<J_L>= (\partial_{\epsilon_1} - \partial_{\epsilon_2})log Z
\eeq
which in the semiclassical limit behaves as
\beq
<J>\propto \frac{(\epsilon_1 - \epsilon_2){\cal{F}}}{(\epsilon_1 \epsilon_2)^2} +...
\eeq
Hence the density of the angular momentum  at weak field
and $\epsilon_1 = - \epsilon_2 = \epsilon$ reads as
\beq
<J_L>\propto \epsilon^{-1}\cal{F}
\eeq
which indicates the instability of the SYM vacuum in the external graviphoton RR 1-form field.
A little bit surprisingly the prepotential itself is proportional
to the density of the angular momentum.

To complete this Section note that having in mind the finite effective
volume in the $\Omega$ background we can write for the "vacuum state" the relation
\beq
<E>_{vac}\propto <J_L>_{vac}
\eeq
which implies that it  effectively behaves as the system
of rotators in the external field. In what follows we shall suggest
the degrees of freedom which could serve as such rotators.

\section{The toy example}

Consider the simplest example
of the  QED in the external electric field $E$. The theory is unstable with respect to the
charged pairs creation. The tunneling  can be
described via the first quantized picture
in terms of the classical
particle trajectories  in the Euclidean space-time \cite{affleck}. The
trajectories are just the circles in the $x,t_E$ plane whose radii
are fixed by the extremization of the simple effective action
\beq
S_{eff}= 2\pi R m - E\pi R^2
\eeq
The coordinate $x$ is selected by the direction of the
external field.
This semiclassical motion in the Euclidean space-time
can be interpreted as the motion in the constant magnetic field.
Similarly the creation of the generic dyons in the
external field can be described in terms of the closed
classical trajectories. The radius of the trajectory
is fixed by the external field moreover the configuration
is unstable. The negative mode in the spectrum
of the fluctuations  implies the
bounce interpretation of such Euclidean solution.

The probability of the pair production in the unit time per unit volume is obtained
from the action calculated  at the bounce solution
\beq
w\propto \exp(-S_{eff}(R_{cr})) \propto \exp(-\frac{m^2}{eE})
\eeq
Since the classical trajectory is the circle with the fixed radius
it implies  the nontrivial angular   momentum $<J>$ from the $R^4$
Euclidean viewpoint in the $x,t_E$ plane.
The density of the loop trajectories in the $t_E,x$ plane corresponds to the angular momentum
density from the $R^4$ viewpoint. Since we are interested in the
derivative of the prepotential with respect to the external field
let us examine the similar expression in QED
\beq
\frac{\partial S_{eff}}{\partial E}\propto <R^2>
\eeq
that is the area inside the semiclassical trajectory.
In the semiclassical approximation $<R^2>\propto R_{cr}^2 \propto E^{-2}$
and therefore diverges at weak field. This behavior reflects the
instability of the system.

The above picture can be generalized for the multiple pair
production problem when the multiple Euclidean bounces have to be taken into
account. Contrary to the single pair case the generic probability
involves the account of the interaction between the bounces which makes the
problem more complicated. The density of the bounces at the 2-dimensional
plane selects the phase  corresponding to "gas" or the "liquid"  state of the
bounce rotators system.

To make the link with ${\cal N}=2$ SYM case let us represent
the pair production in the brane terms
\cite{gss}. The bounce corresponding to the generic $(p,q)$ dyon production
is the tubular $(p,q)$ string  stretched between two D3 branes in IIB setup. In the field
theory limit the deformation of the cylinder due to the finite string tension
is neglected.  In what follows we shall consider very similar
configuration of the tubular D2 branes between the parallel D4 branes in IIA setup.

The brane viewpoint allows to discuss the tunneling problem from the
viewpoint within worldvolume theory of the created brane.
Recall that the Schwinger -type process involves two steps - the tunneling described
by Euclidean bounce
and the Minkowski evolution upon the analytic continuation at the turning
point. For instance, cutting the string cylinder at the turning point we get the disc worldsheet
which represents the open string nucleated in the Minkowski space-time.
The function describing the nucleated branes can be considered as the
Hartle-Hawking wave function $\Psi_{HH}$ in  the worldvolume  theory
\beq
Z(a)\propto \Psi_{HH}(a)
\eeq
where $a$ is some boundary characteristics of the emerging object
and $Z(a)$ corresponds to the probability rate of the brane creation.
Typically the argument of the wave function can be size, metric or some boundary holonomy.
In the case under consideration
the argument of the Hartle-Hawking wave function $\Psi_{HH}$ can be
identified with the boundary length and therefore
$\Psi_{HH}(L)$ plays the role of the "conformal block". We shall
comment on this interpretation later on.
Somewhat related discussion can be found in \cite{hh}.

The Hartle-Hawking interpretation of the
Nekrasov partition function in the semiclassical
approximation is possible as well. To this aim
the prepotential
${\cal{F}}(a)$ has to be interpreted as the action of
some Hamiltonian system. This is true indeed for the
Hamiltonian system with the phase space $(a,a_D)$
\cite{lns,gmmm} due to the set of relations
\beq
a_D=\frac{\partial {\cal{F}}(a)}{\partial a}, \qquad
H(a,a_D)=\frac{\partial {\cal{F}}(a)}{\partial \tau}
\eeq
where $a,a_D$ are the conventional variables defining
the central charge in ${\cal N}=2$ SYM theory \cite{sw}.
Hence prepotential defines the semiclassical
Whitham wave function in the particular polarization
and admits the interpretation as the HH function of the
non-perturbatively created object. Indeed being
the integral over the $A$ cycle the coordinate  variable $a$ could
be treated as the proper boundary data argument of the HH wave function.

\section{On Dyonic instantons and spontaneous vacuum gravimagnetization}
In the previous Section we have explained the analogy
between the effective actions in QED in the constant
background and ${\cal N}=2$ SYM in $\Omega$ background. The Euclidean
bounce is responsible for the pair production and
just these loops of the electrons or generic dyons
in the $R^4$ Euclidean space-time provide
the microscopic description of the angular momentum
which is captured by the derivatives with respect
to the external electric or magnetic field. The induced
angular momentum in the $R^4$ Euclidean space-time
is due to the Euclidean dyonic loops.

The natural question concerns the possible microscopic
description of the angular  momentum in the $\Omega$
background  in terms of some bounces
in the Euclidean space similar to QED.
Since in the $D=5$ SYM one-loop derivation of the Seiberg-Witten
prepotential the sum in the loop goes over the states with
instanton and electric charges \cite{ln} the
dyonic instantons  are the relevant objects to think about.
Similar  states in $\Omega$-background can be thought of
as the blown instantons in the
graviphoton field. The D0 branes get expanded into the
tubular D2 branes
with the density of the D0 branes on their worldvolumes.
In the T-dual picture the dyonic instanton gets represented by
the D1-helix.

The D0 branes are charged with respect to the graviphoton field
and prevent D2 branes from collapsing to a point.
Hence in the $\Omega$ background
we have a kind of bounce configurations in  Euclidean
space-time. Note that somewhat related picture has been developed
in \cite{matsuura} where the Nekrasov partition function
was derived in terms of the blowing of  instantons in the proper
background into the $D2-\bar{D2}$ state.

Let us emphasize that we necessarily
should clarify the origin of the induced angular momentum in the absence
of the external field since the Seiberg-Witten prepotential
which we have related to the density of the angular momentum
can be obtained in the $\epsilon_{1,2}=0$ case as well.
Hopefully contrary to the conventional QED case
the phenomena of the spontaneous gravimagnetization
can happen that is generation of the angular momentum
in $R^4$ without the graviphoton field. In what
follows we shall briefly discuss the solution responsible
for this phenomena. Instead of the external field the
angular momentum stabilizes the extended brane solution in the
Euclidean space-time.

The dyonic instantons can be considered
as the instantons (D0 branes) with attached fundamental strings \cite{dyonic, town,dyonic2}.
Due to the Myers effect the D0+fundamental string state  gets blown
into the tubular D2 branes with the electric field  and the
instanton charge. That is in the $D=4$ space-time
the point-like instanton becomes the circular dyonic
loop which carries the topological charge due to D0 density
and electric field due to the fundamental string.

More precisely the D0 branes are localized at the D4 branes
stretched between two NS5 branes in the standard Hanany-Witten type
brane geometry and their world-line are extended along the coordinate $x_6$
transverse to the NS5 branes. The fundamental string connects two parallel
D4 branes with the geometry $R^4\times I$ where $D=5$ gauge theory lives on.
The dyonic instantons
are solutions to the conventional $D=5$ equation of motion
\beq
F_{\mu\nu}= *F_{\mu\nu},\qquad D_{\mu}\phi =E_{\mu},\qquad D_{0}\phi =0
\eeq
where $\phi$ is the scalar field and $E_{\mu}$ is electric
field in $R^4$.
The shape of the  loop can be arbitrary \cite{town}
however the extremization of the angular momentum yields the
circular  loop. The dyonic instanton is the
BPS state and keeps 1/2 of the initial SUSY. The BPS
formula yields for its mass
\beq
M= \frac{4\pi ^2}{g^2} |I| + |vQ_e|
\eeq
where $|I|$ is the instanton charge and $v$ is the vev
of the scalar. The BPS-ness is provided
by the combined effect of  the scalar,
electric field and   the "running" of the instantons
in the loop.
Note that  the symmetries supported by the
dyonic instantons, that is  diagonal $SU(2)$ from
left rotations and R-symmetry, coincide with the symmetries
kept by the $\Omega$ background.

Let us describe the dynamical quantum numbers of solution
in some details. To this aim it is useful to discuss the
worldvolume Lagrangian of the tubular D2 brane of constant radius
$R$ in flat Euclidean space-time
\beq
L= - \sqrt{R^2(1-E^2) +B^2}
\eeq
where $E,B$ are the worldvolume electric and magnetic fields.
The corresponding canonical momentum reads as
\beq
\Pi =\frac{\partial L}{\partial E}= \frac{R^2 E}{\sqrt{R^2(1-E^2) +B^2}}
\eeq
and Hamiltonian density is
\beq
{\cal H}= R^{-1}\sqrt{(\Pi^2 +R^2)(B^2 + R^2)}
\eeq
There are evident integrals
\beq
Q_F=\frac{1}{2\pi}\oint d\phi \Pi , \qquad Q_0=\frac{1}{2\pi}\oint d\phi B
\eeq
corresponding to the F1 and D0 conserved charges per unit length carried
by the D2 tube.

The tension of the tube is
\beq
T=Q_F=\frac{1}{2\pi}\oint d\phi {\cal H}
\eeq
which equals at the solution to the equation of motion to
\beq
T=|Q_F| + |Q_0|
\eeq
This formula implies that the D2 brane tension does not contribute
hence the cross section of the tube behaves as the "tensionless" object
which explains the arbitrary form of the tube cross section. We have
crossed electric and magnetic fields on the tube that is the Poynting
vector does not vanish and yields the angular component
\beq
M_{\phi}=\Pi B
\eeq
Hence we get the non-vanishing angular momentum of 
dyonic instanton per unit length
\beq
J= Q_F Q_0
\eeq
directed along the axis of the cylinder.

Remind that
the dyonic instanton is extended along $x_6$ hence
looks as closed loop in $R^4$.
The solution carries the angular momentum  which has been identified
in \cite{town}.  
We can calculate the angular momentum exactly on the cross-section
on the supertube \cite{town}
which reads as
\beq
L=\oint d s(x_3\frac{\partial x_4}{\partial s} -
x_4\frac{\partial x_3}{\partial s})
\eeq
It is this angular momentum which provides the stabilization of the
radius of the tubular D2 brane stretched between two D4 branes.

What physics the dyonic instanton represents in the four-dimensional
space-time? The solution is defined in $R^4\times I$ space-time that
is it presumably corresponds to some tunneling process in the Minkowski space.
There are a few arguments supporting such  interpretation. First
note that the dyonic instanton
can be generalized to the multiple circular
tubular D2 branes \cite{emparan}. The solution has zero net
R-R D2 brane charge however there is non-vanishing
dipole four-form field strengths
\beq
G_4=dC_3-dB_2\wedge C_1
\eeq
and 3-form dipole moment is proportional to the angular momentum. Hence
there could be the process of the dropping out the angular momentum
and correspondingly the $G_4$ field via the shell repulsion \cite{emparan}.

In the Euclidean space-time one could search for the negative mode
which upon the analytic continuation would provide  instability
in the Minkowski space. Such negative mode responsible
for the expanding of the radius of the solution has been
found for the large radii in \cite{emparan}. Similar
negative mode in the supertube worldvolume theory
for the large radius has been also found in the Goedel metrics
\cite{drukker}.
Hence at least at large radius of the dyonic instanton
fixed by its angular momentum there is  negative
mode in $R^4$ which supports the bounce interpretation.
Remark that usually the instability reflects the attempt
of the system to screen the  external field like in
Schwinger mechanism. In the dyonic instanton case
the background also involves    the
angular momentum density hence the emerging objects
should have the angular momentum to screen the
background one as well.

Assuming that we are in the region where the negative
mode exists the fate of the solution upon the analytic
continuation can be questioned. The most subtle point
concerns the fate of the topological charge. Indeed
since the solution is just  blown up instanton it carries
the topological charge. What happens with it upon the
analytic continuation from $R^4$ to $R^{3,1}$? Naively we get
the dressed monopole-antimonopole pair hence the initial instanton charge
has to be somehow "divided" between them. Since from Minkowski
viewpoint the instanton corresponds to the tunneling between
two states with the different Chern numbers we can conjecture
that two "dressed monopoles" with the fractional instanton
numbers are involved. Such states are familiar in the theory
with the compact dimensions. Certainly this issue deserves for
further investigation.

\section{Angular momentum in the Liouville theory}
\subsection{The logarithmic operators}
According to the AGT correspondence \cite{agt} the Nekrasov partition function coincides
with the conformal block in the Liouville theory.
Hence we could inspect the
derivatives with respect to the equivariant parameters
at the Liouville side.
Of course we would like to gain some new insights from
this formal procedure. The main goal is to identify
the group of rotation of $R^4$ in Euclidean space
or Lorentz group in the Minkowski version in the
Liouville theory. The naive guess could be that the algebra
generated by screening operators should play the key
role. We shall not provide the complete answer
but a few findings are quite inspiring.

The parameters $\epsilon_i$ enter conformal weights of the
operators involved into conformal block
\beq
\alpha_i= Q/2 + a_{i}
\eeq
and the
central charge
\beq
c= 1 +6Q^2,\qquad Q = b + b^{-1},\qquad b^2 =\frac{\epsilon_1}{\epsilon_2}
\eeq
The derivative
amounts to the substitution of the one
of the vertex operators by the operator
\beq
\tilde{V_{\alpha}}=\phi \exp{\alpha \phi}
\eeq
which is typical logarithmic operator in the CFT. Hence
formally the insertion of the U(1) angular momentum in
$R^4$ corresponds to the insertion of the Liouville field
$\phi$ at the marked points.

Similar expression can be derived in the semiclassical approximation
where the Liouville correlator $\Phi(z_i)$ can be expressed in terms of the
classical Liouville action  calculated at the solutions to the equation
of motion with the prescribed behavior at the insertion points $z_i$
\beq
\Phi(z_i)\propto \exp(\frac{1}{b^2}S_{cl}(z_i)
\eeq
Hence the variation with respect to the equivariant parameters
gets reduced to the variation of the classical action with respect
to the moduli of the solutions. It is convenient to use
the symplectic form on the moduli space of the
solutions to the classical equation of motion in the Liouville theory
\cite{seminara}.
It involves besides the terms term corresponding to
the insertion positions $z_i$ the relevant terms at
each  point
\beq
\omega \propto \sum_{i} d\phi(z_i)\wedge d\alpha_i
\eeq
Therefore we can use this canonically conjugated pair to express
the derivative with respect to the weights
in terms of the values of the Liouville field
\beq
\frac{\partial S}{\partial \alpha_i}\propto \phi(z_i)
\eeq
in agreement with the formal  arguments.

\subsection{Adding the surface operator}

Let us make some comments concerning the effect of the
surface operators on the vacuum
gravimagnetization. Remind that the surface operators
correspond to the D2 branes filling the two-dimensional
sub-manifold of $R^4$ which can be considered from the four-dimensional
viewpoint as the worldsheet  of the magnetic  string.
When both $\epsilon_{1,2}$ parameters are switched on to some extend
one can interpret this background as the condensate of the
magnetic strings. On the gauge theory side the insertion of the
surface operator oriented in some plane in $R^4$ is described
in the semiclassical weak field limit as
\beq
Z_{Nek,sur}= \exp(\frac{{\cal F}}{\epsilon_1 \epsilon_2} + \frac{{\cal W}}{\epsilon_1} +\dots)
\eeq
hence the wave function of the D2 brane reads as
\beq
\Psi_{sur}\propto \exp(\frac{{\cal W}}{\epsilon_1})
\eeq

The derivative of the partition function with the insertion
of the surface operator yields in the semiclassical
limit the additional contribution into
the angular momentum in $R^4$
\beq
\delta <J>\propto {\cal W}(a,z)
\eeq
where coordinate $z$ corresponds to $x_6$ coordinate
of the D2 brane. Equivalently the    ${\cal W}$ has
the interpretation of the twisted superpotential in the
worldvolume theory on D2 brane or the Yang-Yang function
in the  integrable systems \cite{ns}.

Let us turn to the Liouville side where the surface
operator corresponds to the insertion of the
$V_{1,2}$ primary operator \cite{gukov,drukker}. The
contribution of the surface operator
into  gravimagnetization corresponds
to the insertion of the logarithmic primary field
\beq
\tilde{V}_{1,2}= \phi V_{1,2}
\eeq
into the Liouville correlator.
It was found in \cite{zam} that
logarithmic primaries in the semiclassical limit obey
the following relations
\beq
D_{m}\bar{D}_{m}\tilde{V}_{1,m} =A_m V(1,-m)
\eeq
where $A_m$ are numerical constants and
\beq
D_m = \partial^m + d_m
\eeq
where $d_m$ involves the energy stress-tensor
in the Liouville theory and its derivatives.

The operators  $\tilde{V}_{12}$ and $V_{1,-2}$ form the
logarithmic pair of the operators with the same
conformal dimension.
The natural question concerns the interpretation of the
object created by the operator $\tilde{V}_{1,2}$ that is
dressing of  the surface operator.
Some analogy comes from the minimal string theory. In that case there
are two types of the branes in the Liouville theory - ZZ and FZZT branes \cite{zz}.
ZZ branes correspond to the D-instantons  localized in the
Liouville zero mode direction and get condensed.  On the other hand
the FZZT branes are extended in the Liouville coordinate
and  correspond to the surface operators. It was observed
\cite{shih} that the instanton ZZ brane can be interpreted as the
superposition of FZZT brane and anti-brane system.
\beq
|n,m>_{ZZ}= |n,m>_{FZZT} - |n,-m>_{FZZT}
\eeq
We can speculate that
the realization of ZZ brane in terms of the pair of FZZT
branes is  analogous to the blowing up of the instanton.

An interesting interpretation also emerges from the Hamiltonian
viewpoint. The surface operators provide the degrees
of freedom for the particular Hamiltonian system. One of the
equivariant parameters in the $D=4$ SYM theory
can be identified with the
Planck constant in this Hamiltonian system \cite{ns}. 
The same interpretation can be derived in the worldvolume theory
on the surface operator \cite{geras}.
Hence we
have to differentiate  the surface operator wave function with respect
to the Planck constant. From the quantum mechanical viewpoint
the corresponding angular momentum in $R^4$ plays the role
of a kind of the R-symmetry generator. Probably such
picture is related to the Parisi-Sourlas type approach
to the classical dynamics where the effective R-symmetry
can be naturally defined in terms of auxiliary ${\cal N}=2$ SUSY.

\subsection{On the "Liouville type" representation of the effective action}

 To complete this Section let us discuss as byproduct the "Liouville type" representation
of the conventional $D=4$ effective action in the background of constant electromagnetic
field.
To start with let us remind the first quantized representation
for the effective actions for the charged particle of mass $m$
\beq
Z= \sum_{path}\exp(-m L(C) + i\Phi(C)) W(C)
\eeq
where $\Phi(C)$ is the so called spin factor
expressed in terms of the trajectories in $CP^3$  and $W(C)$ is
the Wilson loop along the contour $C$. The potential need
for the Liouville mode to appear is the restoration of the
reparametrization invariance in the summation over
the contours. One
could expect that the integral over the reparametrization
of the contour appears in some form which is familiar
in the stringy calculations of the Wilson loops.

However more convenient form of the effective
action involves the integral over the Schwinger parameter .
In the external self-dual field the one-loop
effective action reads as
\beq
S_{one-loop}= \int \frac{ds}{s} \frac{e^{ism^2}}{(\sinh esE)^2}
\label{one-loop}
\eeq
where we assume that $E=\pm H$. Let us emphasize that
the effective action in QED is not holomorphic object
hence we would like to represent (\ref{one-loop})
as a kind of "correlator" in the Liouville-type theory that
is integrated product of the chiral conformal blocks
over the intermediate conformal weights. Therefore
the question is weather the integral over the Schwinger
parameter can be treated as the integral over the
intermediate  conformal
dimensions in the Liouville theory.

First note that the Schwinger parameter
can be treated as the radial coordinate in the AdS space
\cite{gopa,gorly}. On the other hand the radial coordinate
in AdS geometry can be considered as the zero mode of the
Liouville field \cite{polyakov} that is Schwinger parameter
is related with the Liouville zero mode.
The
correspondence can be made more precise. To the aim
let us use the observation from \cite{gorly} that the
integrand can be represented in terms of the wave function
of the $SL(2,R)$ 2-dimensional YM theory on the disc which
yields the solutions to the $AdS_2$ gravity and depends
on the boundary holonomy as the argument. The boundary
$SL(2,R)$ holonomy can be expressed in terms of the
geodesic length
\beq
tr_{1/2}P\oint _{C} A= 2 cosh\frac{l(C)}{2}
\eeq
which in the case under consideration is proportional
to $eEs$. Hence the Schwinger parameter
is proportional to the geodesic boundary length.

Given  these arguments let us remind that the
boundary lengths $l_i$ are the natural coordinates
on the Teichmueller phase space. It is useful to
consider the coherent
state representation  on the
Teichmueller phase space  in the Kahler polarization
(see \cite{tesh} for the review).
The Liouville conformal block has an interesting
interpretation as the wave function in the quantum
Teichmueller theory in the coherent representation
which is eigenfunction of the length operator \cite{tesh}
\beq
\Psi^{Tei}_{L}(q) = \Psi^{Liov}_{L}(q)
\eeq
Here $L$ is the length variable in the Teichmueller
theory
while simultaneously it parameterizes the weights in the Liouville theory
via identification
\beq
\alpha= Q/2 + \frac{iL}{4\pi b}
\eeq

Hence we see the clear analogy between the Liouville
representation of the ${\cal N}=2$ SYM in the $\Omega$-background
and the QED in the constant external field. The integration
over the dimension of the intermediate state in the conformal
block corresponds to the integration over the Schwinger
parameter in QED. Let us emphasize once again that the holomorphic prepotential
in ${\cal N}=2$ theory corresponds to the conformal block in Liouville
theory while the QED effective action plays the role of the
correlator. Moreover we have mentioned that $s$ has the interpretation
of the zero mode of the Liouville field therefore to some extent
the quantized values of $s_k$ in the imaginary part calculation
correspond to the positions
of Liouville ZZ branes \cite{zz} which are localized at this direction.

\section{Discussion}
In this Letter we have made the first step towards
the investigation  of the
${\cal N}=2$ SYM vacuum response functions in the graviphoton
background focusing on
the simplest  magnetization-like quantity. It is evident
that more complicated response functions, say the
graviconductivity, are interesting as well.

It was argued that
the vacuum state is absolutely unstable with respect to the
generation of the angular momentum in the Euclidean $R^4$ space-time
at arbitrarily weak
graviphoton field. Moreover at the
semiclassical approximation the Seiberg-Witten prepotential
itself is proportional to the angular momentum density.
Our primary goal was to get some new information
concerning SYM vacuum state hence we have tempted to
recognize the microscopic degrees of freedom in $R^4$
responsible for the induced angular momentum. Since
the instantons themselves can  not be the proper degree of freedom
we have looked at the dyonic instantons which are carriers
of the angular momentum.

The analogy with the non-perturbative Schwinger
monopole/dyons pair production
turns out to be fruitful. The process is described by the
bounces in the Euclidean
$R^4$ space-time which are the circular Wilson loop
in the electric case and the circular t'Hooft loop in the
magnetic case. One can naturally assign to these loops the
angular momentum in $R^4$. The dyonic instantons fit
with this picture being  the monopole loop in $R^4$ with the
additional electric and topological charges.

The key point concerns the stabilization mechanism of the
Euclidean bounce solution. We have argued that in the
graviphoton background the bounce stabilization is supported
by the D0 charge density of the tubular D2 brane. However
since the Seiberg-Witten prepotential can be derived
without any graviphoton background another source of the
stabilization has to be presented. The dyonic instanton
indeed provides such mechanism via the nontrivial angular
momentum. Hence we arrive at the interesting picture
of the Euclidean $R^4$ vacuum populated by the dyonic loops
carrying the topological charge density. These loops
emerge as the result of the spontaneous gravimagnetization
of the instanton medium.

It would be interesting to analyze
the possibility of the similar picture of
Euclidean vacuum  populated by the
dyonic instantons in the theories with  less amount of SUSY.
The mechanism of stabilization of the closed curves from
shrinking via the "Poynting angular momentum" seems
to work in much more generic setting. In particular
the QCD-like theory could support the flavored
dyonic instantons stabilized in the similar manner.
The role of the dyonic instantons in the monopole 
condensation upon the perturbation of the ${\cal N}=2$ 
has to be investigated.

The subtle point concerns the
interpretation of medium with the constant density
of   dyonic loops from the
Minkowski viewpoint. In the conventional
Schwinger mechanism  the account of the multiple
bounces corresponds to the sub-leading
factors in the production rate however we do not know
precisely how the medium of the dyonic instantons
could describe the vacuum instability. In general
when considering the multiple bounces the effect of their
interaction has to be taken into account. Moreover
the analytic continuation from the Euclidean to the
Minkowski space appears to be the nontrivial problem.
Indeed when considering the single bounce the selection
of the continuation surface corresponding to the turning
points is a simple issue. However in the case of the multiple
interacting bounces the choice of the proper continuation
surface is complicated problem.

Having in mind the Euclidean
bounce picture  the prepotential itself describes
the semiclassical Hartle-Hawking wave function
of the creating object.
This is consistent with the interpretation of the
prepotential as the logarithm of the Whitham wave
function in the Hamiltonian picture. It would be interesting
to realize the proper meaning of the attractor equations
in our approach.

Using
the AGT relation we commented on the angular momentum issue
in terms of the Liouville conformal blocks and argued
that  the logarithmic operators
in the Liouville theory are relevant. The logarithmic
Liouville theory generically involves the infinite
number of degenerate operators and the role of the
Zamolodchikov higher equations of motion has to be clarified.

It it worth noting that the dyonic  instanton medium in $R^4$
has evident similarities with the Quantum Hall system
hence the methods and concepts developed there could be of some
use in the SYM context. We can not exclude that
the system of strongly interacting dyonic loops
could form a kind of QHE-like vacuum droplet. In this case
the subtle transition from the Euclidean  to the Minkowski
picture would be more simple. Note that the analogy
between the instability of the Goedel metrics and
the motion in the magnetic field has been mentioned in
\cite{drukker2,rey}. In particular the issue of CTC has been
mapped into the problem of the closeness of the
trajectories in the external field.

We also can speculate that
the discussion above has something to do with the Regge
calculus. Indeed we have argued that
dyonic instantons are natural objects populating
Yang-Mills $R^4$ Euclidean vacuum
whose stabilization is supported by the angular momentum.
This issue has some resemblance  with the representation of the
reggeons in the brane terms \cite{gor}. We hope to discuss
these issues elsewhere.

I would like to thank E. Gorsky, V. Mikhailov, A. Mironov,
N. Nekrasov and M. Voloshin for the useful discussions.
I am grateful to FTPI at University of Minnesota and IPhT at Saclay
where the part of the work has been carried out for  hospitality and support.
The work  was supported in part by the grants
RFBR-09-02-00308 and CRDF -  RUP2-2961-MO-09.

\end{document}